\documentclass{emulateapj}

\newcommand{\msun}{{\rm M}_{\odot}}
\newcommand{\rsun}{{\rm R}_{\odot}}
\newcommand{\lsun}{{\rm L}_{\odot}}

\newcommand{\mjup}{{\rm M}_{\rm Jup}}

\newcommand{\kms}{\rm km\ s^{-1}}
\newcommand{\mps}{\rm m\ s^{-1}}

\newcommand{\days}{\rm days}

\newcommand{\spitzer}{{\it Spitzer}}

\begin{document}

\shorttitle{NLTT~41135}
\shortauthors{Irwin et al.}

\title{NLTT 41135: a field M-dwarf + brown dwarf eclipsing binary in a
  triple system, discovered by the MEarth observatory}

\author{Jonathan~Irwin, Lars~Buchhave\altaffilmark{1}, Zachory~K.~Berta,
  David~Charbonneau, David~W.~Latham, Christopher~J.~Burke,
  Gilbert~A.~Esquerdo, Mark~E.~Everett, Matthew~J.~Holman and Philip~Nutzman}
\affil{Harvard-Smithsonian Center for Astrophysics, 60 Garden St.,
  Cambridge, MA 02138, USA}
\email{jirwin -at- cfa.harvard.edu}

\altaffiltext{1}{Current address: Niels Bohr Institute, Copenhagen University, DK-2100 Copenhagen, Denmark}

\author{Perry~Berlind, Michael~L.~Calkins and Emilio~E.~Falco}
\affil{Fred Lawrence Whipple Observatory, Smithsonian Astrophysical Observatory, 670 Mount Hopkins Road,
  Amado, AZ 85645, USA}

\author{Joshua~N.~Winn}
\affil{Department of Physics, and Kavli Institute for Astrophysics and Space
Research, Massachusetts Institute of Technology, Cambridge, MA 02139, USA}

\author{John~A.~Johnson\altaffilmark{2} and J.~Zachary~Gazak}
\affil{Institute for Astronomy, University of Hawaii, Honolulu, HI
  96822, USA}

\altaffiltext{2}{Current address: Department of Astrophysics,
  California Institute of Technology, MC 249-17, Pasadena, CA 91125,
  USA}

\begin{abstract}
We report the discovery of an eclipsing companion to NLTT~41135, a
nearby M5 dwarf that was already known to have a wider, slightly more
massive common proper motion companion, NLTT~41136, at $2\farcs4$
separation.  Analysis of combined-light and radial velocity curves of
the system indicates that NLTT~41135B is a $31-34 \pm 3\mjup$ brown
dwarf (where the range depends on the unknown metallicity of the host
star) on a circular orbit.  The visual M-dwarf pair appears to be
physically bound, so the system forms a hierarchical triple, with
masses approximately in the ratio $8:6:1$.  The eclipses are grazing,
preventing an unambiguous measurement of the secondary radius, but
follow-up observations of the secondary eclipse (e.g. with the James
Webb Space Telescope) could permit measurements of the surface
brightness ratio between the two objects, and thus place constraints
on models of brown dwarfs.
\end{abstract}


\section{Introduction}
\label{intro_sect}

The formation mechanism of brown dwarfs, star-like objects whose masses
are too low to sustain hydrogen fusion in their interiors, has remained
a theoretical puzzle ever since they were first detected.  They pose a
problem for standard star formation theories relying on gravitational
collapse of molecular cloud material because their masses are
substantially smaller than the typical Jeans mass in such a cloud.

Although considerable theoretical and observational effort has been
expended, there is still no clear consensus on how brown dwarfs form.
Popular theoretical ideas can be divided into three major sub-groups.
The first relies on finding some mechanism to produce a sufficiently
low-mass core {\it ab initio}, while circumventing the Jeans mass
problem.  This can be done, for example, by turbulent fragmentation of
cloud material (e.g. \citealt{padoan2004}), fragmentation and
subsequent collapse of molecular cloud cores (e.g. \citealt{boss2002})
or erosion of cores by ionizing radiation from nearby massive stars
(e.g. \citealt{whitworth2004}).  The second posits that brown dwarfs
form as part of multiple systems (which collapse from a single large
overdensity, thus circumventing the Jeans mass problem), and are then
ejected before they can accrete sufficient mass to become normal
hydrogen burning stars (e.g. \citealt{reipurth2001}).  Finally, disk
fragmentation mechanisms (e.g. \citealt{rice2003};
\citealt{stamatellos2009}) are proposed, by which stars and brown
dwarfs form by gravitational instability.  Similar mechanisms have been
proposed for planet formation (e.g. \citealt{boss2006}), and
\citet{kratter2010} note that disk fragmentation in the outer regions
of protoplanetary disks naturally leads to the formation of brown
dwarfs.

Multiple systems offer an important clue as to the nature of the star
formation process and could give an insight into the problem of brown
dwarf formation.  Observations indicate that while multiple systems
are extremely common for solar type stars, with approximately
$65\%$ of systems found to harbor one or more components
(e.g. \citealt{duquennoy1991}), the multiplicity fraction appears to
decline through M spectral types to the brown dwarf domain, with only
$42\%$ of field M-dwarf systems being multiple
(e.g. \citealt{fischer1992}).  The fraction could be as low as
$10\%$ for late-M and L-dwarfs, although there is some disagreement in
the literature regarding corrections for the effects of survey biases,
particularly in binary semimajor axis (see \citealt{burgasser2007} and
references therein for a review).  One formation mechanism that
naturally generates a low multiplicity fraction for brown dwarfs is
the ejection hypothesis mentioned above \citep{reipurth2001} since it
relies on dynamical interactions, which would likely disrupt most
binary brown dwarf systems.  However, this mechanism has difficulty
explaining other observational evidence, such as the existence of
disks around brown dwarfs (e.g. \citealt{jay2003}), which would also
be disrupted by dynamical interactions.

In addition, binary systems have long been a cornerstone in our
understanding of the fundamental properties of stars because gravity
provides one of the few model-independent means to precisely measure
the mass of a star for comparison with other fundamental properties,
thus allowing us to test stellar evolution theory.  Double-lined
eclipsing binaries (EBs) are particularly useful in this regard,
where combined analysis of light curves and radial velocity
information can determine masses and radii for both stars to very high
precision (\citealt{torres2010}; see also the classic review by
\citealt{andersen1991}).

Results of recent dynamical mass measurements for brown dwarf binaries
indicate that the there may be substantial discrepancies between the
predictions of evolutionary models and the observed masses and
luminosities of the objects, by factors of up to $2-3$.  There
has been some debate in the literature as to the nature of these
discrepancies, with the earliest example finding the luminosities were
overpredicted for a given mass (\citealt{close2005}; but this may have
been due to an error in the spectral type, see \citealt{close2007}),
whereas other studies found they are underpredicted
\citep{dupuy2009a}, or indicate no discrepancy
(e.g. \citealt{liu2008}; and \citealt{dupuy2009b} for an intriguing
object at the L/T transition).  By making measurements of an
unprecedentedly large sample of objects, \citet{konopacky2010}
conclude that the discrepancies are a function of spectral type, with
late-M to mid-L systems having luminosities overpredicted by the
theoretical models, and one T-type system having a luminosity which is
underpredicted, indicating that the sense of the discrepancies may
reverse moving to later spectral types.  Furthermore, the evolutionary
and atmospheric models may indeed be inconsistent with one another
(e.g. \citealt{dupuy2009a}; \citealt{konopacky2010}), indicating
systematic errors in one or both classes of models.

The only well-characterized double lined brown dwarf EB is perhaps
even more puzzling, showing a reversal in effective temperatures of
the two objects, with the more massive brown dwarf being cooler
\citep{stassun2006,stassun2007}.  This is thought to be due to
magnetic activity in this extremely young ($\sim 1-2\ {\rm Myr}$)
system \citep{chabrier2007}.  What is clear is that even more
dynamical measurements of brown dwarfs are needed, to place these
results on a firmer footing and to probe a wider range of parameter
space.

In this work we present the detection of an eclipsing brown dwarf
orbiting the lower-mass member of a known field M-dwarf visual double,
\object[NLTT 41136]{NLTT~41136} and \object[NLTT 41135]{41135}, which
we show are highly likely to be physically associated.  The
observation of eclipses constrains the orbital inclination, and thus
allows the secondary mass to be ascertained through radial velocity
measurements with essentially no $\sin i$ ambiguity.  The properties
of the parent M-dwarf can also be constrained from the eclipse
measurements.  Such an object represents an opportunity to constrain
formation and evolutionary models of both M-dwarfs and brown dwarfs.

\section{Observations and data reduction}
\label{obs_sect}

\subsection{MEarth photometry}
\label{mearth_obs_sect}

Eclipses in NLTT~41135 were initially detected from data taken during
2009 February to May (inclusive) as part of routine operation of the
MEarth observatory, a system designed primarily to search for
transiting super-Earth exoplanets orbiting around the nearest $2000$
mid to late M-dwarfs in the northern hemisphere
(\citealt{nutzman2008}; \citealt{irwin2009}).  Exposure times on each
field observed by MEarth are tailored to achieve sensitivity to the
same planet size for the assumed parameters of each target star,
and were $151{\rm s}$ for the field containing NLTT~41135.

Data were reduced using the standard MEarth reduction pipeline, which
is at present nearly identical to the Monitor project pipeline
described in \citet{irwin2007}.  We used an aperture radius of $10\
{\rm pixels}$ ($7\farcs6$); note that the MEarth telescopes were
operated slightly out of focus in the 2008 September to 2009 July
observing season to increase the number of photons that can be
gathered before saturation of the detector for our brightest targets.

This star was identified as a high-priority candidate from a
box-fitting least squares (BLS) transit search (following closely the
methods of \citealt{burke2006}), performed on de-trended, median
filtered light curves.  We used a $3$-day median filter
\citep{aigrain2004} to remove stellar variability, followed by a trend
filtering algorithm (TFA; \citealt{kovacs2005}) based on searching for
optimal linear combinations of comparison star light curves to best
remove any systematic effects in the target light curve.  The
photometry indicated a period of approximately $2.8\ {\rm days}$, and
a $2\%$ eclipse depth, with no secondary eclipses visible in the
MEarth data.  We immediately switched to a follow-up mode after the
initial detection, observing the object at the highest possible
cadence (approximately $3$ minutes including overheads) during the
night of UT 2009 May 25 when an eclipse was predicted to occur.  This
was confirmed with high significance, although any measurements of the
eclipse shape were severely corrupted by the target crossing the
meridian during the eclipse, which necessitates a temporary halt in
the observations and rotation of the telescope optics and detector
system through $180^\circ$ relative to the sky with our German
Equatorial Mounts.  In practice, the telescope guiding takes some time
to recover after this, and photometry is corrupted during the recovery
period.  Therefore while we were able to confirm the reality of the
event, the MEarth data are not useful in determining the eclipse shape
or physical properties of the object.

Due to the close proximity of NLTT~41135 and 41136, these stars are
unresolved in the MEarth observations, and indeed are also unresolved
or improperly resolved in many well-known literature sources for
measurements of bright stars.  It was initially not clear which member
of the M-dwarf pair was undergoing eclipses, so we pursued photometric
(resolved light curves) and spectroscopic (radial velocity) methods to
determine the identity of the eclipsing star.  The spectroscopic
method gave the earliest indication that NLTT 41135 was responsible
for the eclipses.

\begin{figure}
\centering
\includegraphics[angle=0,width=2.8in]{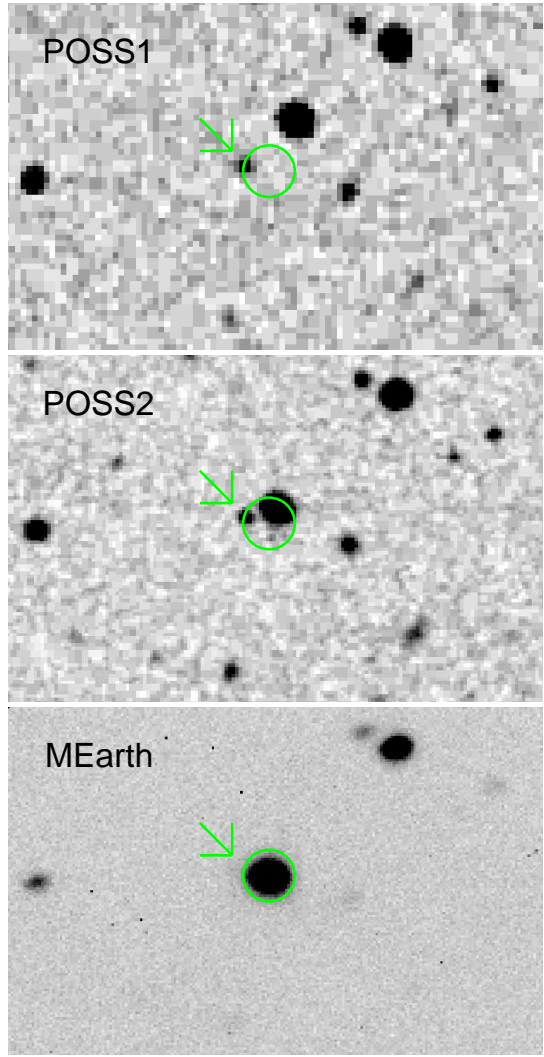}
\caption{Images of NLTT~41135 and 41136 centered on the position as
  measured from the MEarth data.  The circle shows the approximate
  position and size of the $7\farcs6$ (radius) photometric aperture
  used to derive our light curves, and the arrow points to the
  position of the fainter star discussed in the text.  Data are from
  the first and second epoch Palomar sky surveys as provided by the
  Digitized Sky Survey (DSS; top and center panels), and the MEarth
  master image (bottom panel).  The two stars are unresolved, although
  elongation of the combined image is visible in the POSS-2 (center)
  panel.  The approximate epochs of the images are $1954.4$ (POSS-1),
  $1993.2$ (POSS-2), and $2009.1$ (MEarth).  All three panels have the
  same center, scale and alignment on-sky, with north up and east to
  the left, covering approximately $2\farcm5$ in the horizontal
  direction.}
\label{third_light_images}
\end{figure}

The high proper motion of the pair allows us to constrain the
contribution of any additional background stars in the photometric
aperture that are not co-moving with NLTT~41135 and 41136, using
previous  epochs of imaging.  We show in Figure
\ref{third_light_images} a series of three images centered on the
position of the photometric aperture used in the MEarth
images.  These show that there is a fainter star located approximately
$7$ arcsec to the east and $1$ arcsec to the north of the blended pair
at the MEarth epoch.  Using the APM on-line sky catalog\footnote{\tt
  http://www.ast.cam.ac.uk/\~{}apmcat/}, this star
is $4$ magnitudes fainter than the blended image of NLTT~41135 and
41136 in the POSS-1 red (E) plate, and $2.3$ magnitudes fainter in the 
blue (O) plate.  While this star is at the edge of the MEarth
photometric aperture and thus will affect the eclipse depths
measured from these data, we have ensured that this star was
excluded from the apertures used to analyze the follow-up data
described in the subsequent sections, and its fainter magnitude and
bluer color should ensure any contamination in the far-red bandpasses
used in this work is negligible.

Our MEarth light curves show variations out of eclipse.  These have a
peak-to-peak amplitude of $\simeq 0.04\ {\rm mag}$, and rms scatter of
$0.012\ {\rm mag}$.  There is no clear period, and it is not clear
which member of the pair is responsible for the variations.  Our
photometry is corrupted somewhat by the partially-resolved nature of
the pair, so we regard this measurement as an upper limit to the
actual variability level, noting that the absence of a clear period
may be indicative of some or all of the measured variation being
spurious.

\subsection{UH 2.2m $z$-band resolved follow-up photometry}
\label{uh_sect}

In order to determine the flux ratio of the M-dwarf pair, and hence
the intrinsic flux decrement during eclipse, we obtained a single,
resolved image of the system using the Orthogonal Parallel Transfer
Imaging Camera (OPTIC; \citealt{howell2003}; \citealt{tonry2005}) on
the University of Hawaii 2.2m telescope, during the night of UT 2009
June 9.  This instrument uses two $2048 \times 4096$ ``orthogonal
transfer'' CCDs (e.g. \citealt{tonry1997}) which allow parallel
transfers of the charge through the image area in both the $x$ and $y$
directions.  The pixel scale is $0\farcs14$/pix on the 2.2m telescope,
giving a field of view of $9\farcm3 \times 9\farcm3$.  An exposure
time of $10\ {\rm s}$ was used in good seeing conditions through the
$z$ filter.  The Heliocentric Julian Date of mid-exposure was
$2454991.910257$, corresponding to an orbital phase of $0.23$ (close
to quadrature) for the NLTT~41135 system (see \S \ref{ephem_sect}).

This single image was bias-subtracted using the overscan region, and
flat-fielded before using our standard source detection software to
obtain positions and aperture photometry for both stars in the pair.
We measured an angular separation of $2\farcs40$, and a flux ratio of
$l_{41136,z}/l_{41135,z} \equiv L_3 = 2.1 \pm 0.2$, where we have
assumed a conservative error based on multiple measurements of the
image made using different sized apertures.  This quantity will later
be used in the analysis of unresolved light curves of the system to
extract the parameters of the eclipsing binary, NLTT~41135.  In this
context it is usually referred to as ``third light'', and shall
hereafter be labeled with the symbol $L_3$ for consistency with the
usual nomenclature in the eclipsing binary literature.

It is important to consider the possible influence of stellar
variability on the value of $L_3$.  Using the results stated in \S
\ref{mearth_obs_sect}, in a similar passband, the most pessimistic
assumption is to assign all of the measured variation to NLTT~41135,
giving a contribution to the fractional uncertainty in $L_3$ of $4\%$
(rms).  Our assumed uncertainty of $\pm 0.2$ (a fractional uncertainty
of $10\%$) in $L_3$ is already sufficient to account for this, and the
additional contribution from variability is probably negligible.

An attempt to obtain a resolved light curve was made on the night of
UT 2009 June 17.  Exposure times of $25\ {\rm s}$ were used through
the $z$ filter.  We used the orthogonal parallel transfer capability
of the OPTIC detector to shift charge during the exposure in an
attempt to perform a tip-tilt correction of the incoming wavefronts to
improve the FWHM of the images.  However, due to technical problems
this instead caused severe corruption of the point spread functions,
which limits the utility of the resulting light curve.  In addition,
during the eclipse the guiding lock was lost which caused a short gap
in the phase coverage.  Nonetheless, we were able to extract resolved
photometry of the two stars by performing standard aperture photometry
on the resulting images.  An aperture radius of $4\ {\rm pixels}$
($0\farcs56$) was found to give the best compromise between flux
losses from the aperture and cross-contamination of the point spread
functions of the two stars.  We used NLTT~41136 as the comparison star
to extract differential photometry of NLTT~41135, which has the
additional advantage of canceling much of the cross-contaminated flux
between the two objects and hence providing a more accurate
measurement of the true eclipse depth.  This light curve is given in
Table \ref{optic_z_lc_table}.

\begin{deluxetable*}{lrrrrrrr}
\tabletypesize{\normalsize}
\tablecaption{\label{optic_z_lc_table} UH 2.2m resolved $z$-band light curve.}
\tablecolumns{8}

\tablehead{\colhead{HJD\tablenotemark{a}} & \colhead{Differential $z$} & 
    \colhead{Error\tablenotemark{b}} &
    \colhead{$\Delta m$\tablenotemark{c}} & \colhead{FWHM\tablenotemark{d}} & \colhead{Airmass} &
    \colhead{$x$\tablenotemark{e}} & \colhead{$y$\tablenotemark{e}} \\
& & & & \colhead{(pix)} & & \colhead{(pix)} & \colhead{(pix)}
}

\startdata
$2454999.853406$ &$-0.0502$ &$0.0039$ &$-0.567$ &$6.36$ &$1.03609$ &$482.569$ &$3423.312$ \\
$2454999.854022$ &$-0.0358$ &$0.0064$ &$-0.474$ &$2.95$ &$1.03604$ &$474.552$ &$3421.937$ \\
$2454999.854670$ &$ 0.0192$ &$0.0043$ &$-0.371$ &$6.02$ &$1.03601$ &$477.566$ &$3419.809$ \\
$2454999.855283$ &$ 0.0048$ &$0.0037$ &$-0.088$ &$5.88$ &$1.03599$ &$475.421$ &$3421.553$ \\
$2454999.855896$ &$ 0.0261$ &$0.0040$ &$-0.103$ &$6.32$ &$1.03600$ &$469.411$ &$3419.771$ \\
\enddata

\tablenotetext{a}{Heliocentric Julian Date of mid-exposure.  All HJD
  values reported in this paper are in the UTC time-system.}
\tablenotetext{b}{Estimated using a standard CCD noise model,
  including contributions from Poisson noise in the stellar counts, sky
  noise, readout noise and errors in the sky background estimation.}
\tablenotetext{c}{Correction to the frame magnitude zero-point applied
  by the differential photometry procedure.  More negative numbers
  indicate greater losses.  Please note that this has already been
  applied to the ``differential $z$'' column and is provided only for
  reference (e.g. distinguishing frames with large losses due to cloud).}
\tablenotetext{d}{Median FWHM of the stellar images on the frame.  The
  plate scale was $0\farcs14/{\rm pix}$.}
\tablenotetext{e}{$x$ and $y$ pixel coordinates on the CCD image,
  derived using a standard intensity-weighted moments analysis.}

\tablecomments{Table \ref{optic_z_lc_table} is published in its
  entirety in the electronic edition of the Astrophysical Journal.  A
  portion is shown here for guidance regarding its form and content.}

\end{deluxetable*}

\subsection{FLWO $1.2\ {\rm m}$ $z$-band follow-up photometry}
\label{kepcam_sect}

Observations centered around the primary eclipse of UT 2009 June 20
were obtained using the KeplerCam instrument on the Fred Lawrence
Whipple Observatory (FLWO) $1.2\ {\rm m}$ telescope. We used the
standard binning $2 \times 2$ readout mode, since the pixel scale of
$0\farcs34$ per unbinned pixel significantly oversamples the typical
seeing at FLWO.  The resulting scale was $0\farcs67$ per summed pixel.
We used the $z$ filter and an exposure time of $120\ {\rm s}$.  A
total of $86$ observations were taken, starting approximately $1\ {\rm
hour}$ before first contact and finishing $0.5\ {\rm hours}$ after
last contact to sample the out-of-eclipse portions of the light curve,
thus allowing the eclipse depth to be properly measured.

These photometric data were reduced using the same pipeline as
described in \S \ref{mearth_obs_sect}.  The FWHM of the stellar images
was approximately $2\farcs5$, so we used an aperture radius of $8$
binned pixels, corresponding to $5\farcs4$ on-sky, to extract aperture
photometry of the combined light of NLTT~41135 and NLTT~41136,
reproduced in Table \ref{z_lc_table}.  Our attempts at PSF fitting
photometry on the images to extract individual light curves did not
yield useful results.

\begin{deluxetable*}{lrrrrrrr}
\tabletypesize{\normalsize}
\tablecaption{\label{z_lc_table} FLWO 1.2m $z$-band light curve.}
\tablecolumns{8}

\tablehead{\colhead{HJD\tablenotemark{a}} & \colhead{Differential $z$} & 
    \colhead{Error\tablenotemark{b}} &
    \colhead{$\Delta m$\tablenotemark{c}} & \colhead{FWHM\tablenotemark{d}} & \colhead{Airmass} &
    \colhead{$x$\tablenotemark{e}} & \colhead{$y$\tablenotemark{e}} \\
& & & & \colhead{(pix)} & & \colhead{(pix)} & \colhead{(pix)}
}

\startdata
$2455002.740052$ &$-0.0000$ &$0.0012$ &$ 0.013$ &$3.91$ &$1.12786$ &$497.620$ &$524.552$ \\
$2455002.741672$ &$ 0.0025$ &$0.0012$ &$-0.006$ &$4.00$ &$1.12904$ &$497.732$ &$524.679$ \\
$2455002.743188$ &$ 0.0000$ &$0.0013$ &$-0.019$ &$4.20$ &$1.13026$ &$497.800$ &$524.485$ \\
$2455002.744751$ &$ 0.0013$ &$0.0012$ &$ 0.007$ &$3.87$ &$1.13162$ &$497.716$ &$524.532$ \\
$2455002.746302$ &$-0.0001$ &$0.0013$ &$ 0.016$ &$3.85$ &$1.13307$ &$497.629$ &$524.695$ \\
\enddata

\tablenotetext{a}{Heliocentric Julian Date of mid-exposure.  All HJD
  values reported in this paper are in the UTC time-system.}
\tablenotetext{b}{Estimated using a standard CCD noise model,
  including contributions from Poisson noise in the stellar counts, sky
  noise, readout noise and errors in the sky background estimation.}
\tablenotetext{c}{Correction to the frame magnitude zero-point applied
  by the differential photometry procedure.  More negative numbers
  indicate greater losses.  Please note that this has already been
  applied to the ``differential $z$'' column and is provided only for
  reference (e.g. distinguishing frames with large losses due to cloud).}
\tablenotetext{d}{Median FWHM of the stellar images on the frame.  The
  plate scale was $0\farcs67/{\rm pix}$.}
\tablenotetext{e}{$x$ and $y$ pixel coordinates on the CCD image,
  derived using a standard intensity-weighted moments analysis.}

\tablecomments{Table \ref{z_lc_table} is published in its
  entirety in the electronic edition of the Astrophysical Journal.  A
  portion is shown here for guidance regarding its form and content.}

\end{deluxetable*}

\subsection{Sloan Digital Sky Survey photometry}

Our target lies within the survey area for the Sloan Digital Sky
Survey (SDSS).  In Data Release 7 (DR7; \citealt{sdssdr7}), the object
is resolved in the SDSS images, but appears to be blended with a false
galaxy detection in the band-merged catalogs, so some of the
magnitudes may be unreliable.  We therefore re-derive resolved
magnitudes for the two stars from the reduced images.

The relevant {\tt fpC} ``corrected frames'' were retrieved from the
SDSS archive.  We used the $r$-band image for source detection, since
this had the smallest FWHM of $\simeq 2.8\ {\rm pixels}$.  The
positions derived therefrom were then used to perform aperture
photometry on the $g$, $r$ and $z$ band images, using an aperture
radius of $2.5\ {\rm pixels}$ ($1\farcs0$) to avoid overlap between
the apertures placed on the two components of the visual binary.  The
$i$ image appears to be corrupted for NLTT~41136, so we did not
attempt to derive photometry from it, and the $u$ band image showed
very low counts for the M-dwarf pair, as expected, so this was also
omitted, and is likely of limited utility in any case due to a
well-documented red leak issue with the SDSS $u$ filter affecting
measurements of very red targets.

The final calibrated magnitudes are given in Table \ref{photparams}.
We estimate uncertainties in these measurements by comparing our
measurements of other stars in the image with the SDSS DR7 photometry,
and allowing for cross-contaminated flux between the apertures placed
on our targets.  The photometric errors in $g$ are large in part due
to the greater FWHM of the stellar images in this band compared to $r$
and $z$, necessitating a larger aperture correction.  The orbital
phase of these measurements for the NLTT~41135 system was
approximately $0.34$ (see \S \ref{ephem_sect}), and the measured
$z$-band flux ratio is $2.03$, which is consistent with the value
reported from the OPTIC data in \S \ref{uh_sect}.

Our measurements in $g$ and $z$ for our target stars are
systematically brighter than those reported in the SDSS DR7 catalog,
which we suspect to be due to de-blending of the false ``galaxy''
detection from the stellar images.  The images of our target do not
appear to be saturated in the SDSS data.

\subsection{FLWO $1.5\ {\rm m}$ TRES spectroscopy}
\label{tres_sect}

Spectroscopic observations of both stars were obtained using the TRES
fiber-fed \'echelle spectrograph on the FLWO $1.5\ {\rm m}$
Tillinghast reflector.  We used the medium fiber ($2\farcs3$ projected
diameter), yielding a resolving power of $R \simeq 44\,000$.

We obtained two epochs on each star in the visual binary close to
the predicted times of quadratures from the photometric ephemeris, in
order to search for large-amplitude radial velocity variability.  One
hour exposures were obtained in June 2009 over a four day period.

The radial velocities were extracted using a custom-built pipeline to
rectify and cross correlate the spectra.  This pipeline was identical
to the one used for the Nordic Optical Telescope data and will be
described in the next section.

The two radial velocities of NLTT~41135 from TRES showed variations in
phase with the prediction from the photometric ephemeris, giving an
initial estimate of the RV semi-amplitude of $K = 12.9 \pm 0.5\
\kms$.  We cross correlated the spectra of the two stars and
subtracted the orbital motion of the low-mass companion of
NLTT~41135. If the pair of stars are physically bound, this relative
velocity should be constant over short (few day) timescales and close
to zero.  We found a velocity difference of $860 \pm 130\ \mps$ over
$4\ {\rm days}$, which is consistent with the pair being physically
bound, within the errors.  These results encouraged us to obtain high
signal-to-noise high-resolution spectra to characterize the RV
variations of NLTT~41135.

We note that NLTT~41135 shows the H$\alpha$ line in emission, whereas
NLTT~41136 does not, and instead there is a hint of an absorption
feature at this wavelength in our low signal-to-noise spectra.  This
is reasonably consistent with expectation for a field-age M-dwarf
system given the rapid increase in the observed H$\alpha$ activity
fraction around M4 to M5 spectral types (e.g. \citealt{west2004}), but
it may also be indicative of enhanced activity on NLTT~41135 due to
tidal effects from the close companion causing it to rotate more
rapidly.

\subsection{NOT/FIES spectroscopy}
\label{fies_sect}

Precise radial velocities of NLTT~41135 were obtained using the
FIbre-fed \'Echelle Spectrograph (FIES) on the 2.5m Nordic Optical
Telescope (NOT) at La Palma, Spain. We used the medium resolution
fiber ($1\farcs3$ projected diameter) with a resolving power of  $R
\simeq 46\,000$ giving a wavelength coverage of $\simeq 3600-7400\
\rm{\AA}$. We obtained seven high-resolution, high signal-to-noise
spectra of NLTT~41135 during a seven night run in August 2009.

The spectra were rectified and cross correlated using a custom built
pipeline designed to provide precise radial velocities for \'echelle
spectrographs. The procedures are described in more detail in Buchhave
el al. (2010, in preparation) and will be described briefly below.

In order to effectively remove cosmic rays, each observation was split
into three separate exposures, enabling us to combine the raw images
using median filtering, removing virtually all cosmic rays.  We use a
flat field to trace the \'echelle orders and to correct the pixel to
pixel variations in CCD response, then extract one-dimensional spectra
using the ``optimal extraction'' algorithm of \citet{hewett1985} (see
also \citealt{horne1986}). The scattered light in the two-dimensional
raw image is determined and removed by masking out the signal in the
\'echelle orders and fitting the inter-order scattered light flux with
a two-dimensional polynomial.  The FIES sky fiber was broken at the
time these observations were taken, so sky subtraction could not be
performed.  Unfortunately, this means we are also unable to measure
equivalent widths, e.g. to characterize the activity of the star.

Thorium argon (ThAr) calibration images were obtained through the
science fiber before and after each stellar observation.  The two
calibration images are combined to form the basis for the fiducial
wavelength calibration. FIES and TRES are not vacuum spectrographs and
are only temperature controlled to 1/10 of a degree. Consequently, the
radial velocity errors are dominated by shifts due to pressure,
humidity and temperature variations. In order to successfully remove
these large variations ($> 1.5\ \kms$), it is critical that the
ThAr light travels along the same optical path as the stellar light
and thus acts as an effective proxy to remove these variations. We have
therefore chosen to sandwich the stellar exposure with two ThAr frames
instead of using the simultaneous ThAr fiber, which may not exactly
describe the induced shifts in the science fiber and can also lead to
bleeding of ThAr light into the science spectrum from the strong argon
lines, especially at redder wavelengths. The pairs of ThAr exposures
are co-added to improve the signal to noise ratio, and centers of the
ThAr lines are found by fitting a Gaussian function to the line
profiles and a two-dimensional fifth order Legendre polynomial is used
to describe the wavelength solution.

Once the spectra have been extracted, a cross correlation is performed
order by order.  The spectra are interpolated to a common oversampled
log wavelength scale, high and low pass filtered and apodized. The
orders are cross correlated using a Fast Fourier Transform (FFT) and
the cross correlation functions (CCFs) for all the orders co-added and
fit with a Gaussian function to determine the RV.  For the FIES
spectroscopy typically 10 orders (of a total of 78) yielded usable
cross-correlations, where the majority of the orders we did not use
are in the blue and had very poor signal to noise ratios.  For
TRES, 8-10 orders were used. Uncertainties were estimated by fitting
Gaussians to the individual orders' cross correlation functions, and
taking the error in the mean of all the orders.  The template used in
this procedure was constructed from the target spectra, initially
determining the RVs by cross-correlating against the best
signal-to-noise individual spectrum of the target, and then shifting
and adding all of the spectra to make a single high signal-to-noise
stacked template, and repeating the correlations against this to
produce the final RVs.

The radial velocity measurements of NLTT~41135 are reported in Table
\ref{rv_table}.

\begin{deluxetable}{lrrr}
\tabletypesize{\normalsize}
\tablecaption{\label{rv_table} Barycentric radial velocity measurements of NLTT~41135.}
\tablecolumns{4}

\tablehead{\colhead{HJD\tablenotemark{a}} & \colhead{$v$\tablenotemark{b}} & \colhead{$\sigma_v$} & \colhead{Seeing\tablenotemark{c}} \\
& \colhead{($\mps$)} & \colhead{($\mps$)} & \colhead{(arcsec)}
}

\startdata
$2455048.392461$ &$22098.6$ &$31.5$ &$1.0$ \\
$2455049.390174$ &$  -83.2$ &$27.4$ &$0.9$ \\
$2455050.385098$ &$ 6408.9$ &$48.8$ &$0.9$ \\
$2455051.384396$ &$21245.8$ &$25.4$ &$0.8$ \\
$2455052.383913$ &$-2062.3$ &$33.4$ &$0.7$ \\
$2455053.381672$ &$ 9464.6$ &$38.2$ &$1.1$ \\
$2455054.373863$ &$19808.2$ &$52.7$ &$1.8$ \\
\enddata

\tablenotetext{a}{Heliocentric Julian Date of mid-exposure.  All HJD
  values reported in this paper are in the UTC time-system.}
\tablenotetext{b}{Relative radial velocity.  The zero-point is
  arbitrary and corresponds to the original reference spectrum (HJD
  $2455049.39$).  An estimate of the actual $\gamma$ velocity is
  reported in the text and given in Table \ref{rvparams}.}
\tablenotetext{c}{Seeing estimates for a wavelength of $500\ {\rm nm}$
  and an airmass of 1.0 from the Differential Image Motion Monitor
  (DIMM) operated by the Isaac Newton Group on the same site.  See {\tt
  http://www.ing.iac.es/Astronomy/development/seeing/} for more 
  information on this instrument.}
\end{deluxetable}

Since our radial velocities are measured with respect to the target
itself, the zero-point is arbitrary.  We make a separate estimate of
the systemic velocity itself (usually called $\gamma$ in eclipsing
binary studies, and we follow the same notation in this work) using
the H$\alpha$ emission lines in NLTT 41135.  These were fit with
Gaussian functions, and the resulting spectroscopic orbit was analyzed
in the same way as described in \S \ref{rv_sect} to determine the
$\gamma$ velocity and its error, reported in Table \ref{rvparams}.

We do not see evidence for significant rotational broadening in the
observed cross-correlation functions, where the formal FWHM of the
cross-correlation peak was $\approx 12\ \kms$.  This implies a FWHM of
$8.5\ \kms$ in the individual spectra as we used the target as
its own cross-correlation template.  Formally, given the stated
resolving power, this would imply a contribution of $\approx 5\ \kms$
from the star, although the uncertainty in this value is large and we
do not believe the small difference in FWHM is significant in
practice, as other sources can inflate the measured cross-correlation
function width.  Unfortunately, we lack observations of suitable
slowly-rotating M-dwarf templates to make a robust measurement of the
rotational broadening.  However, the lack of a coherent, periodic
modulation in the MEarth photometry (see \S \ref{mearth_obs_sect})
also indicates probable low rotation, where $v \sin i = 5\ \kms$ would
imply a rotation period of $\la 2.1\ \days$.

\subsection{Summary of the photometric and astrometric system properties}
\label{ssp_sect}

Table \ref{photparams} summarizes the known system properties, from
the literature (principally the proper motion survey of
\citealt{ls2005} used to select the target stars for the MEarth
survey) and the SDSS photometry.  A source appears in the 2MASS point
source catalog, but we do not report these measurements here since
the photometry is flagged as having a very poor point spread function
fit and may therefore be unreliable.  Comparing the predicted and
measured J-band magnitudes indicates that the source detected in
2MASS is most likely NLTT~41136 only.

\begin{deluxetable}{lrr}
\tabletypesize{\normalsize}
\tablecaption{\label{photparams} Summary of the photometric and
  astrometric properties of NLTT~41135 and NLTT~41136.}
\tablecolumns{3}

\tablehead{\colhead{Parameter} & \colhead{NLTT~41136} & \colhead{NLTT~41135}
}

\startdata
$\alpha_{2000}$\tablenotemark{a,b}    & $15^h46^m04^s.41$             & $15^h46^m04^s.26$ \\
$\delta_{2000}$\tablenotemark{a,b}    & $+04^\circ41\arcmin31\farcs2$ & $+04^\circ41\arcmin30\farcs0$ \\
$\mu_\alpha \cos \delta$\tablenotemark{b}   & \multicolumn{2}{c}{$0\farcs156\ {\rm yr^{-1}}$} \\
$\mu_\delta$\tablenotemark{b}               & \multicolumn{2}{c}{$-0\farcs284\ {\rm yr^{-1}}$} \\
\\
$g$  & $16.261 \pm 0.06$ & $17.422 \pm 0.06$ \\ 
$r$  & $14.927 \pm 0.03$ & $16.036 \pm 0.03$ \\
$z$  & $12.374 \pm 0.03$ & $13.145 \pm 0.03$ \\
\\
Spectral type &M$4.2 \pm 0.5$ &M$5.1 \pm 0.5$ \\
$T_{\rm eff}$\tablenotemark{c} &$3340 \pm 120\ {\rm K}$ &$3230 \pm 130\ {\rm K}$ \\
\\
$d_{\rm phot}$           &$24.2 \pm 11.2\ {\rm pc}$ &$22.7 \pm 6.8\ {\rm pc}$ \\
\enddata

\tablenotetext{a}{Equinox J2000.0, epoch 2000.0.}
\tablenotetext{b}{From \citet{ls2005}.  These authors measure only the
combined proper motion of the pair, and not individual proper motions
for the two stars.  We have been unable to locate individual estimates
in the literature.}
\tablenotetext{c}{We assume a $\pm 100\ {\rm K}$ systematic
  uncertainty in the effective temperatures.}

\end{deluxetable}

\begin{figure}
\centering
\includegraphics[angle=270,width=2.8in]{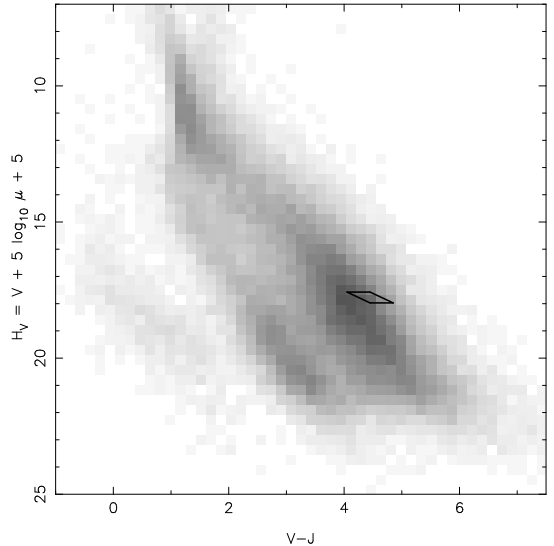}
\caption{$V$ versus $V-J$ reduced proper motion diagram for LSPM
  stars, reproduced from \citet{ls2005}, but plotted as greyscales and
  showing the position of NLTT~41136 by the parallelogram shaped box.
  See the text for a discussion of the uncertainties in determining
  the correct position for this object.}
\label{rpm}
\end{figure}

As a check that these objects are indeed nearby dwarfs, we show in
Figure \ref{rpm} the position of our source on a V versus ${\rm V-J}$
reduced proper motion (hereafter, RPM) diagram, reproduced from
Fig. 30 of \citet{ls2005}.  Only NLTT~41136 is shown since these
authors do not provide estimates of magnitudes for NLTT~41135.  It is
not clear if the source was resolved in the surveys used to gather
their photometry (USNO-B1.0 for V and 2MASS for J), so we indicate the
$0.4\ {\rm mag}$ uncertainty resulting from the possible inclusion of
NLTT~41135 in either, or both, bands by the parallelogram shaped box
on the diagram. Nevertheless, it appears the position of NLTT~41136 is
consistent with that expected for a nearby dwarf in RPM.

We derive photometric estimates of the spectral types for both stars
using the average SDSS colors for M-dwarfs of \citet{west2005},
interpolating between the measured $r - z$ values in their Table 1.
The derived spectral types and their uncertainties are reported in our
Table \ref{photparams}.  Note that these uncertainties are largely
systematic in nature, i.e. the error in relative spectral type between
the two stars is smaller than the errors we report.  We have also
converted the observed spectral types to effective temperature using
the scale of \citet{kenyon1995}.  We assume a systematic uncertainty
of $\pm 100\ {\rm K}$ in the effective temperatures following
\citet{torres2002}.

Finally, photometric estimates of the distance to each star were
derived using the absolute magnitude relations from \citet{west2005},
and comparing to the measured magnitudes.  These distances differ by
$< 1 \sigma$, which, in combination with their apparent common proper
motion, strongly supports our conclusion that the two stars are
physically associated.

\subsection{Ephemeris determination}
\label{ephem_sect}

The small number of radial velocity measurements gathered does not in
practice place useful constraints on the ephemeris.  We therefore
derived this entirely from photometry, using the discovery data (after
applying the filtering detailed in \S \ref{mearth_obs_sect}) and the
times reported in \S \ref{lc_sect} for the two well-sampled eclipses.
These were modeled with a linear ephemeris, fitting the KeplerCam light
curve model from \S \ref{lc_sect} simultaneously to the discovery data
and the two eclipse timings.

We estimate the uncertainties in the ephemeris parameters using the
``residual permutation'' bootstrapping method
(e.g. \citealt{winn2009}) to account for the effects of the correlated
noise (systematics) present in the MEarth data on the eclipse times.
The light curve was first fit adjusting three parameters: the
magnitude zero-point, ephemeris zero-point $t_0$ and period $P$, to
compute best-fitting values for these parameters.  We then proceeded
to fit the model to new light curves generated by taking the original
best-fit and adding in cyclic permutations of the residuals from that
fit.  The final parameter values and their associated uncertainties
were estimated using the median and central $68.3\%$ confidence
intervals of the distributions of the new parameters from these fits,
over all $N - 1$ cyclic permutations of the data (where $N = 993$ is
the number of data points in the MEarth light curve).  In this way,
the correlations in the noise are taken into account in the results.

We find that the times of mid-primary eclipse are given by $t_k = t_0
+ k P$ where $k$ is an integer, with $t_0 = 2455002.80232 \pm 0.00022$
(HJD) and $P = 2.889475 \pm 0.000025\ {\rm days}$.

\section{Radial velocity analysis}
\label{rv_sect}

We fit a single Keplerian orbit model to the observed radial
velocities using a Levenberg-Marquardt non-linear least-squares
method.  Since the M-dwarf host star NLTT~41135 emits most of its
light in the infrared and the spectral coverage of FIES ends at
approximately 7400 \AA, we were able to use only the reddest orders
that were not affected by telluric lines.  Typically $10$ orders were
used in the analysis.  The average internal error for the 30 minute
exposures was $37\ \mps$. We have quadratically added a ``jitter''
term of $17\ \mps$ to the velocities to yield a reduced $\chi^2 =
1.0$.  We therefore estimate that the true error of the velocities is
$\simeq 41\ \mps$.

\begin{figure}
\centering
\includegraphics[angle=0,width=3.2in]{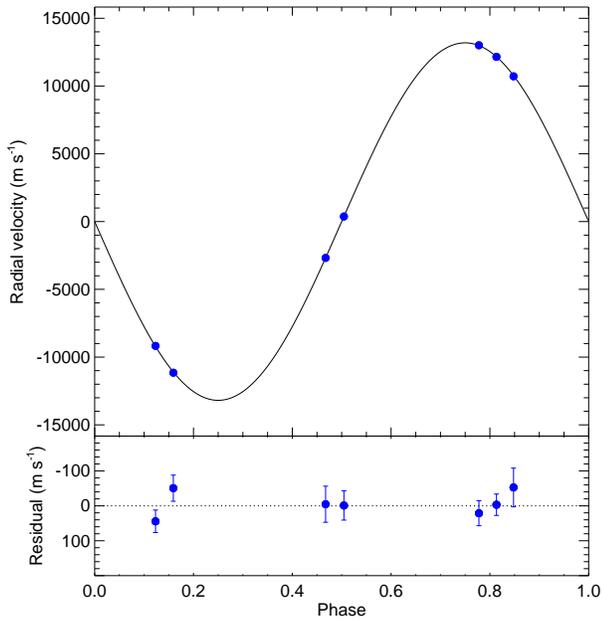}
\caption{Upper panel: Radial velocity measurements from the Nordic
  Optical Telescope as a function of orbital phase (measuring from 0
  at the primary eclipse and in units of the orbital period; for the
  assumed circular orbit the secondary eclipse is at 0.5) with the
  best orbital fit overplotted. The $\gamma$ velocity of the system
  has been subtracted and the fit assumes a circular orbit, fixing the
  ephemeris to that found by the photometry.  The internal error bars
  are plotted, but are not visible due to their small size.  Lower
  panel: Phased residuals of the velocities after subtracting the best
  fit model. The rms variation of the residuals is $39\ \mps$. Note
  that the scale of the lower panel has been expanded to better show
  the residuals.}
\label{phaseplot}
\end{figure}

We fixed the mid-eclipse time and period to the values found from the
photometric analysis of the light curve and assumed a circular orbit,
which leaves two free parameters, namely the velocity amplitude and
the $\gamma$ velocity of the system. The radial velocities and resulting
best-fitting model are shown in Figure \ref{phaseplot}, and the
parameters derived therefrom are presented in Table \ref{rvparams}.
The rms variation of the residuals from the best fit model was $39\
\mps$.

In order to place an upper limit on the eccentricity, we also
carried out fits where the eccentricity and argument of periastron
were allowed to vary (whilst enforcing consistency with the
photometric ephemeris).  This yielded an eccentricity of $e=0.007 \pm
0.005$ and $K = 13.269 \pm 0.062\ \kms$.  We can thus constrain the
eccentricity to be $e < 0.02$ with 99\% confidence.

\begin{deluxetable}{lrl}
\tabletypesize{\normalsize}
\tablecaption{\label{rvparams} Parameters for the radial velocity 
  model of NLTT~41135.}
\tablecolumns{3}

\tablehead{
\colhead{Parameter} & \colhead{Value} &
}

\startdata
$\gamma$ ($\kms$)    &$41.1 \pm 1.2$ \\
$K_1$ ($\kms$)       &$13.189 \pm 0.019$ \\
$e$                  &$< 0.02$ & ($99\%$ confidence) \\
\enddata

\end{deluxetable}

\section{Light curve analysis}
\label{lc_sect}

For detached eclipsing binaries on circular orbits, the radial
velocity (RV) and light curve models are largely independent.  Since
the radial velocity measurements show no evidence for non-zero
eccentricity, we assume a circular orbit for the purposes of analyzing
the light curves.

Since we do not detect secondary eclipses, and the light from the
secondary is expected to be negligible in the optical, with a
predicted z-band luminosity ratio of at most $\simeq 10^{-3}$ between
NLTT~41135B and NLTT~41135A for ages appropriate to an old, field
system (see \S \ref{sec_sect}), we assume a dark secondary for the
purposes of modeling the light curve.  This allows the use of simple
models developed for fitting transiting exoplanet light curves.

We use the analytic transit curves of \citet{mandel2002} to model the
KeplerCam primary eclipse light curve, modeling limb darkening with a
standard quadratic law (e.g. \citealt{claret1990};
\citealt{vanhamme1993}) of the form:
\begin{equation}
{I_\lambda(\mu)\over{I_\lambda(1)}} = 1 - u_1(1-\mu) - u_2(1-\mu)^2
\end{equation}
where $I_\lambda(\mu)$ expresses the monochromatic intensity at
wavelength $\lambda$ for a point on the surface at angle $\theta$ from
the normal, $\mu = \cos \theta$, and $u_1$ and $u_2$ are free
parameters.

We accounted for the dilution of the eclipse depth by the presence of
NLTT~41136 in the photometric apertures using the measured $z$-band
light ratio ($L_3$) from \S \ref{uh_sect}.  Since the light ratio and
the light curves themselves were both observed through $z$ filters,
the error introduced by doing this should be small, and we account for
its effect in the final uncertainties by assuming a conservative error
on $L_3$.

After accounting for the known system ephemeris, the model has five
free parameters: the ratio of component radii $R_2/R_1$, orbital
inclination $i$, the semimajor axis divided by the primary radius,
$a/R_1$, and two limb darkening parameters $u_1$, $u_2$.  In
practice, we introduce three additional parameters: the normalization
(out of eclipse magnitude) $z_0$, a linear term in airmass $k (X-1)$
(where $X$ is airmass) to account for any residual differential
atmospheric extinction effects, and we also allow for a timing offset
$\Delta t$ to account for any error in the ephemeris or deviations
in the eclipse time due to orbital perturbations.

While it is usually possible (with sufficiently high-quality data) to
fit for all three geometric parameters ($R_2/R_1$, $i$ and $a/R_1$) of
a total eclipse, it is important to note that this is not possible for
a grazing eclipse, because the second and third contact points are no
longer seen in the light curve.  In the simple case with no limb
darkening and a dark secondary, this can be understood as follows.
For a total eclipse, it is straightforward to show that the eclipse
depth is determined by the area ratio between the two objects, and is
thus equal to $\left(R_2/R_1\right)^2$.  The eclipse duration, and the
duration (or equivalently, slope) of the partial phases (first to
second contact, and third to fourth contact) are determined by $i$ and
$a/R_1$ (e.g. see \citealt{seager2003} for a detailed derivation).  In
the grazing case, although we still know the eclipse duration and the
slope of the partial phases, we do not know what the level of the
``bottom'' of a total eclipse would be.  Therefore, the parameter
$R_2/R_1$ is essentially undetermined without external information on
the other two.

Since we are primarily interested in $i$ and $a/R_1$, we adopt a weak
Bayesian prior on the most poorly determined parameter, $R_2/R_1$, and
marginalize over this parameter to determine the distribution of
possible values of the other two.  In addition, the grazing light
curve only constrains the limb darkening parameters very weakly, so we
fix these at values appropriate for the $z$ passband from
\citet{claret2004} using the effective temperature from Table
\ref{photparams} and assuming $\log g = 5.0$ and solar metallicity.
The theoretical uncertainty in limb darkening parameters for the $z$
passband should be negligible compared to the intrinsic degeneracies
in the modeling of the grazing eclipse for the present case.

In order to derive parameters and reliable error estimates, we adopt a
variant of the popular Markov Chain Monte Carlo analysis frequently
applied for transiting exoplanet systems (e.g. \citealt{tegmark2004};
\citealt{ford2005}; \citealt{winn2007}).  The majority of these
analyses use the standard Metropolis-Hastings method
\citep{metropolis1953,hastings1970} with Gibbs sampler, which has the
advantage of being extremely simple to implement.  We briefly describe
this method here (as typically used for parameter estimation in
astronomy), and then summarize the modifications made in our
implementation.

Starting from an initial point in parameter space, the
Metropolis-Hastings algorithm takes the most recent set of parameters
and perturbs one or more parameters by a random Gaussian deviate.
Gibbs sampling perturbs one parameter at a time, cycling around the
parameters.  If the perturbed parameter set has a lower $\chi^2$ than
its progenitor, it is accepted as a new point in the chain.  If it has
a larger $\chi^2$, it is accepted with a probability $\exp(-\Delta
\chi^2/2)$.  If it is not accepted, the original point is repeated in
the chain.  The size of the perturbations are usually adjusted by
manual iteration so that $20-30\%$ of the proposed points are
accepted.

We adopt the Adaptive Metropolis algorithm of \citet{haario2001},
which dynamically updates a Gaussian proposal distribution during the
simulation using the empirical covariance matrix, thus largely
eliminating the need for initially tuning the proposal distributions
to obtain the correct acceptance rates.  This is by definition
non-Markovian, but the updates are done in a way that has been shown
to maintain the correct ergodicity properties of the chain.

We used a standard Levenberg-Marquardt algorithm\footnote{\tt
  http://www.ics.forth.gr/\~{}lourakis/levmar/} to provide the initial
parameter and covariance estimates to start the chain, and used the
resulting fit to re-scale the observational errors such that the
reduced $\chi^2$ was equal to unity.

For parameter estimation, we used a chain of $10^7$ points, discarding
the first $20\%$ of these in order to ensure the chain had converged.
The correlation lengths for all parameters were $< 100$ points.  We
enforced the observed luminosity ratio between NLTT~41136 and
NLTT~41135, of $L_3 = 2.1 \pm 0.2$ as discussed in \S \ref{uh_sect}.

As discussed earlier in this section, we have assumed a prior
on the radius ratio, to break the degeneracy between the parameters in
the light curve model.  Without any prior, the allowed values of
$R_2/R_1$ are bounded only below (corresponding to the limiting case
where the eclipse switches from being grazing to total), and are
essentially unbounded above despite $R_2/R_1 \ga 1$ being physically
extremely unlikely.

From the radial velocities, we are left with little doubt that the
secondary lies below the hydrogen burning limit
(e.g. \citealt{chabrier2000} and references therein) and is therefore
a brown dwarf.  The physics of this transition is quite well
understood, and except for deuterium burning, the only source of
energy for these objects is gravitational contraction.  Theoretical
models predict that all objects with masses in this range should have
radii $< 0.11\ \rsun$ after $300\ {\rm Myr}$ \citep{burrows1997},
implying $R_2/R_1 < 0.6$.  The measured radii of OGLE-TR-122b
\citep{pont2005} and CoRoT-3b \citep{deleuil2008} also corroborate
this theoretical argument.  We can place an analogous lower limit,
for example, secondary sizes smaller than the planet Neptune would
require implausibly high densities at the measured secondary mass,
implying $R_2 / R_1 > 0.17$.

We therefore adopt a Gaussian prior, $R_2/R_1 = 0.4 \pm 0.3$, the
value of which was derived by taking the predicted  radius ratio from
the stellar and brown dwarf models given the estimated masses of the
two components, but also encapsulates the argument made in the
previous paragraph.  We assume a conservative error to  account for
the uncertainty in this estimate.  An isotropic prior (i.e. uniform in
$\cos i$) was assumed on the orbital inclination, and  uniform priors
were assumed on $z_0$, $k$ and $R_1 / a$. 

\begin{figure}
\centering
\includegraphics[angle=270,width=3.2in]{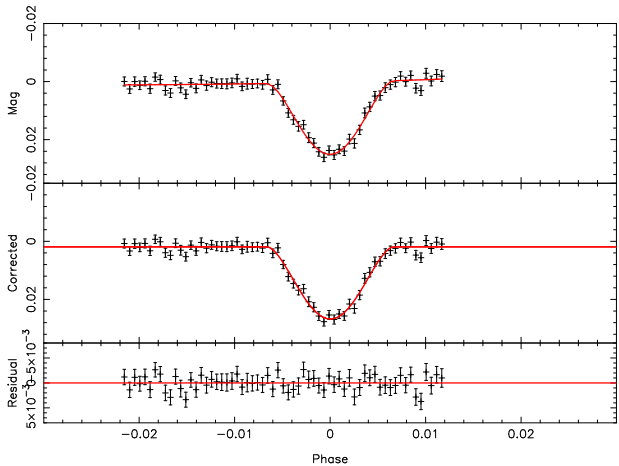}
\caption{Top panel: phase-folded $z$-band KeplerCam light curve with
  our best-fitting light curve model over-plotted.  Center panel:
  as above, with the linear airmass term subtracted out before
  plotting.  Bottom panel: residual after subtracting the model from
  the data.}
\label{lcplot}
\end{figure}

\begin{figure*}
\centering
\includegraphics[angle=270,width=6in]{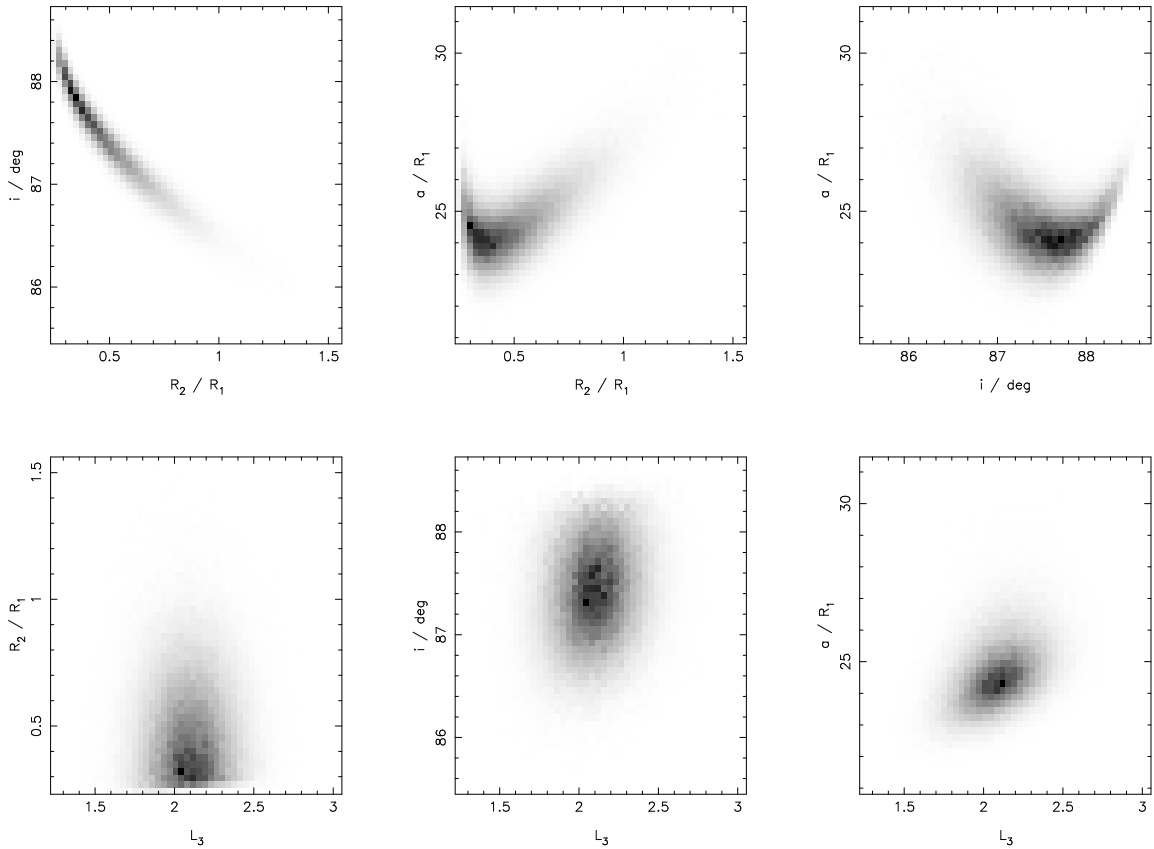}
\caption{Two-dimensional greyscale histograms of the posterior
  probability density functions for selected parameter pairs from our
  Monte Carlo analysis.  The mapping from probability density to the
  intensity of the greyscales was a square root function, to compress
  the dynamic range and allow the tails of the distribution to be more
  readily distinguished.}
\label{lcmodparamsplot}
\end{figure*}

Our final parameter values and their associated uncertainties were
estimated using the median and central $68.3\%$ confidence intervals
of the marginalized posterior probability distributions for each
parameter, and are reported in Table \ref{lcparams}.  Figure
\ref{lcplot} shows the best-fitting model overplotted on the light
curves, and in Figure \ref{lcmodparamsplot} we show representative
two-dimensional distributions of the parameters of the model to
illustrate the discussion of the degeneracies in the light curve
modeling from this section.  It is clear from the table that despite
the inherent degeneracies in the light curve model, we do indeed
obtain acceptable bounds on the inclination and $a/R_1$, especially
recalling that the inclination appears as $\sin i$ when we come to
interpret the radial velocity measurements.

\begin{deluxetable}{lrl}
\tabletypesize{\normalsize}
\tablecaption{\label{lcparams} Parameters for the light curve
  model of NLTT~41135.}
\tablecolumns{3}

\tablehead{
\colhead{Parameter} & \colhead{Value} &
}

\startdata
$t_0$ (HJD)          &$2455002.80232 \pm 0.00022$ \\
$P$ (days)           &$2.889475 \pm 0.000025$     \\
\\
$u_1$                &$0.0169$          &(assumed) \\
$u_2$                &$0.6976$          &(assumed) \\
$L_3$                &$2.10 \pm 0.20$   &(assumed) \\
\\
$\Delta t$ (HJD)     &$0.00006 \pm 0.00023$ \\
$k$ (mag/airmass)    &$-0.0062 \pm 0.0023$ \\
$i$ (deg)            &$87.42^{+0.50}_{-0.51}$ \\
$a / R_1$            &$24.60^{+1.18}_{-0.93}$ \\
$R_2/R_1$            &$0.48^{+0.24}_{-0.15}$ \\
\enddata

\end{deluxetable}

Note that the choice of a linear term in airmass $k (X - 1)$ to
represent the variations in the out-of-eclipse baseline is somewhat
arbitrary.  While it is reasonable to expect atmospheric extinction to
affect the differential photometry at the observed level (especially
given the much redder color of our target relative to the comparison
stars), it is equally possible the observed ``slope'' is due to
stellar variability.  This means in practice that the depth
information used in the model fits is uncertain due to the unknown
contribution of stellar spots.

Assuming a sinusoidal out of eclipse modulation with the same period
as the orbit, the measured value of $k$ would correspond to a
semi-amplitude of $\ga 0.01\ {\rm mag}$, or correspondingly, $\ga
0.005\ {\rm mag}$ if the period was half an orbit (e.g. ellipsoidal
variation).  While the MEarth photometry places limits on the actual
level of out of eclipse variation (see \S \ref{mearth_obs_sect}), and
thus on the possible phases and periods corresponding to the measured
$k$, the unexplained out of eclipse variations seen there may indeed
be consistent with the ``slope'' seen in the KeplerCam data.

\begin{figure}
\centering
\includegraphics[angle=270,width=3.2in]{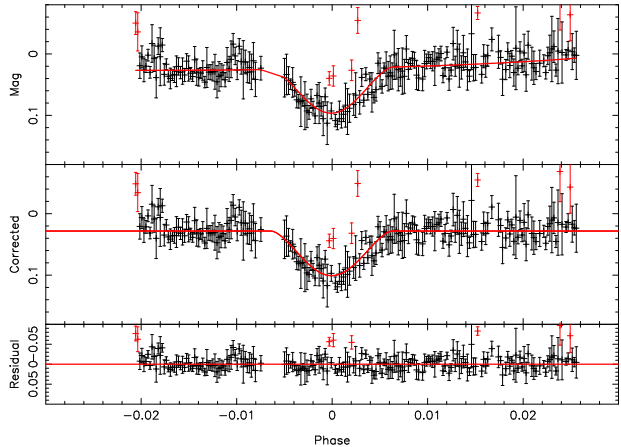}
\caption{Phase-folded resolved $z$-band light curve from
  the OPTIC data with our best-fitting light curve model over-plotted.
  Panels as Figure \ref{lcplot}.  Points that were excluded from the
  fit by our clipping procedure are colored red in the online version
  of the figure.}
\label{uhlcplot}
\end{figure}

We use the resolved light curve from \S \ref{uh_sect} to verify the
parameters we have derived from the KeplerCam curve, shown in Figure
\ref{uhlcplot}.  We re-fit for the three parameters $z_0$, $k$ and
$\Delta t$, and assumed $L_3 = 0$ (since the light curve was
resolved), but all other parameters were held fixed from the KeplerCam
analysis.  The fit was carried out using $3 \sigma$ clipping, in order
to reduce the effect of the large number of photometric outliers on
the fit results, and we re-scaled the observational errors such that
the reduced $\chi^2$ of the out of eclipse parts of the light curve
was equal to unity.  The reduced $\chi^2$ of the full fit (including
the eclipse) was $1.04$.

As an additional check, we have also performed a simultaneous fit of
the KeplerCam and OPTIC light curves.  Separate normalization, $k$ and
$\Delta t$ values were allowed for each of the two curves, and we fit
for a single set of geometric parameters ($k$, $i$, and $a/R_1$) and
for $L_3$ (removing the prior used in the previous analyses and
instead fitting for this parameter).  We used the same clipping and
rescaling of errors carried over from the separate analyses, which
should correct for any differences in the under-estimation of the
observational uncertainties between the two curves and produce the
correct relative weights.

The values of $i$, $R_2/R_1$ and $a/R_1$ from this analysis differ by
$\ll 1 \sigma$ from those given in Table \ref{lcparams}, providing
further verification of the results.  We derive $L_3 = 2.15 \pm 0.16$,
which is consistent with the value assumed in the analysis of the
KeplerCam light curve to $< 1 \sigma$.  The $99\%$ confidence interval
from this fit is $1.80 < L_3 < 2.52$.

Finally, we report the mid-eclipse times from our analysis of the
two well-sampled eclipses from the unresolved KeplerCam and the
resolved OPTIC light curves in Table \ref{minlight}.

\begin{deluxetable}{lrrl}
\tabletypesize{\normalsize}
\tablecaption{\label{minlight} Measured mid-eclipse times for NLTT~41135.}
\tablecolumns{4}

\tablehead{
\colhead{HJD\tablenotemark{a}} & \colhead{$(O - C)$ (s)} &
\colhead{Cycle\tablenotemark{b}} &\colhead{Instrument}
}

\startdata
$2454999.91269$ &$-13 \pm 37$ &$-1$ &OPTIC \\
$2455002.80238$ &$5 \pm 20$   &$0$  &KeplerCam \\
\enddata

\tablenotetext{a}{Heliocentric Julian Date of mid-eclipse.  All HJD
  values reported in this paper are in the UTC time-system.}
\tablenotetext{b}{Cycle number, counting from $0$ at the primary
  eclipse at $t_0$, in units of the orbital period.}

\end{deluxetable}

We now proceed in the next section to interpret the parameters derived
from the radial velocity and light curve modeling in terms of the
physical parameters of the M-dwarf NLTT~41135A and the brown dwarf
NLTT~41135B.

\section{Properties of NLTT~41135A and system physical parameters}
\label{pri_sect}

Using the $\gamma$ velocity from Table \ref{rvparams} and the measured 
positions, proper motions and photometric parallax from Table
\ref{photparams}, we estimate $(U,V,W) = (-53.8 \pm 6.2, -5.0 \pm 3.0,
4.9 \pm 5.9)\ \kms$ using the method of \citet{johnson1987}, where we
adopt the definition that positive $U$ values are away from the
Galactic center.  This places NLTT~41135 in the old Galactic disk
population as defined by \citet{leggett1992}.  While kinematics
provide relatively weak constraints on the stellar ages, these are
typically a few to $10\ {\rm Gyr}$ for old disk stars, and it is thus
highly likely that NLTT~41135 is older than a few Gyr.

In order to derive an estimate of the mass of NLTT~41135A, we adopt
the methodology of \citet{torres2007}, although the available
information in our case is more limited since we do not have
a reliable measurement of the parallax or near-infrared magnitudes.

The quantity $a / R_1$ from the light curve model in \S
\ref{lc_sect} provides a measurement of the mean density of the
primary, $\bar{\rho_1}$ using Newton's version of Kepler's
third law:

\begin{equation}
\left({a\over{R_1}}\right)^3 = {G (M_1 + M_2)\over{R_1^3 \Omega^2}} =
     {G (1 + q)\over{\Omega^2}}\ {4 \pi \bar{\rho_1}\over{3}}
\end{equation}

where $\Omega = 2 \pi / P$ is the angular frequency of the orbit.  We
explicitly note that the secondary mass is not negligible in the
present case.

We use this estimate of the stellar density in conjunction with
stellar evolution models to derive $M_1$.  For this work,
we adopt the tracks of \citet{bcah98}.

It is well-established from studies of eclipsing binary stars 
(e.g. \citealt{lopezmorales2007}; \citealt{chabrier2007}) that stellar
evolutionary models tend to underpredict the radii of M-dwarfs at a
given mass.  The bolometric luminosities from the same models do not
seem to be seriously in error (e.g. \citealt{delfosse2000};
\citealt{torres2002}; \citealt{torres2006}; \citealt{torres2007}).  We
therefore follow \citet{torres2007} and introduce a multiplicative
correction factor $\beta$ to the radii, while at the same time
applying a factor $\beta^{-1/2}$ to the effective temperatures to
conserve the bolometric luminosity.  Studies of eclipsing binaries
(e.g. \citealt{ribas2008}) and the transiting planet host GJ~436
\citep{torres2007} indicate that typical values of $\beta \simeq
1.1$.

In practice, we fit for the observed stellar density and effective
temperature (from Table \ref{photparams}), given a fixed metallicity,
and allowing the age to vary.  We do not use the observed optical
colors as an additional constraint, because these are known to be
poorly-reproduced by the models (e.g. \citealt{bcah98}).  This is
thought to be the result of a missing source of molecular opacity in
the {\sc NextGen} model atmospheres.

The kinematic information presented in this section indicates that
our target is an old, field M-dwarf.  We therefore explore a range of
ages from $1 - 10\ {\rm Gyr}$, noting that this has little effect on
the derived parameters in practice, because M-dwarfs in this age range
evolve relatively slowly.  The metallicity is also unknown, but
typical values for old disk stars are ${\rm [Fe/H]} \simeq -0.5$
(e.g. \citealt{leggett1992}).  We derive parameters for both this
value and for solar metallicity, to give an idea of the likely range
resulting from this source of uncertainty, which does have a
significant effect on the stellar parameters.

Our derived physical parameters for NLTT~41135A, the orbital
parameters, and the inferred mass of NLTT~41135B, are given in Table
\ref{physparams}.  The value of $\beta$ indicates essentially no
inflation of the radius of NLTT~41135A relative to the models, within
the observational errors, but this is not surprising since the
propagated error in the effective temperature measurement yields only
a weak constraint on this parameter, with the value of $1.1$ found in
other studies also being reasonably consistent with the measurements.

\begin{deluxetable}{lll}
\tabletypesize{\normalsize}
\tablecaption{\label{physparams} Derived physical parameters for the
  NLTT~41135 system.}
\tablecolumns{3}

\tablehead{
\colhead{Parameter} & \multicolumn{2}{c}{Value} \\
 & \colhead{${\rm [Fe/H]} = 0.0$} & \colhead{${\rm [Fe/H]} = -0.5$}
}

\startdata
$M_1$ ($\msun$)      & $0.188^{+0.026}_{-0.022}$ & $0.164^{+0.021}_{-0.018}$ \\
$R_1$ ($\rsun$)      & $0.210^{+0.016}_{-0.014}$ & $0.201^{+0.014}_{-0.013}$ \\
$L_1$ ($\lsun$)      & $0.0043^{+0.0012}_{-0.0010}$ & $0.0039^{+0.0012}_{-0.0009}$ \\
$\log g_1$           & $5.062^{+0.033}_{-0.034}$ & $5.072^{+0.033}_{-0.034}$ \\
$\beta$              & $1.021^{+0.057}_{-0.053}$ & $1.079^{+0.059}_{-0.055}$ \\
\\
$a$   ($\rsun$)      & $5.15^{+0.21}_{-0.20}$ & $4.93^{+0.19}_{-0.17}$ \\
$q$                  & $0.171^{+0.008}_{-0.008}$ & $0.180^{+0.008}_{-0.008}$ \\
\\
$M_2$ ($\mjup$)      & $33.7^{+2.8}_{-2.6}$ & $30.9^{+2.4}_{-2.1}$ \\
\enddata

\end{deluxetable}

Given the limited constraints available on the properties of
NLTT~41135A, it is perhaps not surprising that the knowledge of the
mass of the brown dwarf secondary, NLTT~41135B, is presently
limited by the uncertainty in the primary mass.  One of the most
reliable methods to constrain single M-dwarf masses in this range is
to use the mass-luminosity relation, either empirical determinations
such as \citet{delfosse2000} or the models.  We therefore suggest that
the most profitable way to proceed would be to obtain an accurate
trigonometric parallax, and resolved, near-infrared apparent
magnitudes for NLTT~41135 and NLTT~41136.  Of the available passbands,
K would appear to be the logical choice, since the
\citet{delfosse2000} K-band relation is one of the best-constrained
and shows the smallest scatter.  This should allow the uncertainty in
the primary mass to be reduced to $\sim 10\%$, and permit a test of the
models of M-dwarfs in addition to improving the secondary parameters.
We also note that the passband-integrated luminosities of M-dwarfs can
be used to estimate the metallicity (e.g. \citealt{johnson2009}).

Since NLTT~41135 is a member of a presumed visual binary system,
ultimately it may be possible to measure its dynamical mass from the
orbit of the visual binary.  This would allow a model-independent
radius to be derived for NLTT~41135A, permitting a direct test of
M-dwarf evolution models, and would also provide a model-independent
dynamical mass for NLTT~41135B.  We discuss this possibility in \S
\ref{vb_sect}.

\section{Prospects for constraining brown dwarf evolution models}
\label{sec_sect}

While we estimate the mass of NLTT~41135B to $10-15\%$ (limited by the
uncertain metallicity and accuracy of determination of the mass of the
M-dwarf host), the lack of a secondary eclipse and the detection of
only one object in the spectrum mean that this is the only parameter
we can presently determine.

The most straightforward approach to obtaining a second fundamental
parameter for the brown dwarf is to attempt to detect a secondary
eclipse.  The ratio of the primary to secondary eclipse depths in a
given passband is approximately the ratio of central surface
brightnesses of the two objects, $J = L_2 R_1^2 / L_1 R_2^2$, where
all these quantities are measured for the specific passband in
question.  The division by the squared radius ratio makes this a more
favorable prospect than direct detection of light from the secondary
e.g. in a spectrum, and many successful detections of shallow
secondary eclipses have been achieved for exoplanet systems,
especially in the mid-infrared using the \spitzer\ space telescope
(e.g. \citealt{charbonneau2005}; \citealt{deming2005,deming2006}).

Table \ref{secpredict} gives predicted luminosity ratios, surface
brightness ratios, and secondary eclipse depths assuming the value of
$M_2$ from Table \ref{physparams} and using brown dwarf evolution
models to predict the surface brightness of NLTT~41135B in
near-infrared bandpasses.  We use the {\sc cond} models of
\citet{baraffe2003} for the secondary, and the {\sc NextGen} models
discussed in \S \ref{pri_sect} for the primary.  We do not compute
predictions for bluer bandpasses than $z$ due to known problems with
both sets of models, as mentioned earlier (see also \citealt{baraffe2003}
for discussion of problems with the I-band magnitudes in the {\sc
cond} models), but the predicted eclipse depths are even shallower at
these wavelengths due to the rapidly declining flux from the brown
dwarf, so detection here seems unlikely.  Three representative ages,
$1$, $5$ and $10\ {\rm Gyr}$, are shown in the table, with the
kinematic evidence discussed in \S \ref{pri_sect} favoring the old end
of this age range.

\begin{deluxetable}{lrlrrr}
\tabletypesize{\normalsize}
\tablecaption{\label{secpredict} Predicted luminosity ratios, surface
  brightness ratios, and secondary eclipse depths for NLTT~41135 with
  $A_g = 0.0$.}
\tablecolumns{6}

\tablehead{
\colhead{Age} & \colhead{$T_2$} & \colhead{Band} & \colhead{$L_2/L_1$} & \colhead{$J$} &\colhead{$d_2$\tablenotemark{a}} \\
& \colhead{(K)} & & & & \colhead{(\%)}
}

\startdata
$1\ {\rm Gyr}$  &$1067$ &z    &$8.60 \times 10^{-4}$ &$3.94 \times 10^{-3}$ &$0.03$ \\
                &       &J    &$3.51 \times 10^{-3}$ &$1.61 \times 10^{-2}$ &$0.12$ \\
                &       &H    &$1.97 \times 10^{-3}$ &$9.04 \times 10^{-3}$ &$0.068$ \\
                &       &K    &$1.59 \times 10^{-3}$ &$7.29 \times 10^{-3}$ &$0.055$ \\
                &       &L$'$ &$1.04 \times 10^{-2}$ &$4.76 \times 10^{-2}$ &$0.36$ \\
                &       &M    &$1.73 \times 10^{-2}$ &$7.95 \times 10^{-2}$ &$0.6$ \\
\hline
$5\ {\rm Gyr}$  &$643$  &z    &$1.17 \times 10^{-4}$ &$6.24 \times 10^{-4}$ &$0.0047$ \\
                &       &J    &$1.84 \times 10^{-4}$ &$9.81 \times 10^{-4}$ &$0.0074$ \\
                &       &H    &$1.09 \times 10^{-4}$ &$5.82 \times 10^{-4}$ &$0.0044$ \\
                &       &K    &$2.04 \times 10^{-5}$ &$1.09 \times 10^{-4}$ &$0.00082$ \\
                &       &L$'$ &$1.18 \times 10^{-3}$ &$6.31 \times 10^{-3}$ &$0.047$ \\
                &       &M    &$4.26 \times 10^{-3}$ &$2.27 \times 10^{-2}$ &$0.17$ \\
\hline
$10\ {\rm Gyr}$ &$533$  &z    &$4.64 \times 10^{-5}$ &$2.61 \times 10^{-4}$ &$0.002$ \\
                &       &J    &$5.19 \times 10^{-5}$ &$2.92 \times 10^{-4}$ &$0.0022$ \\
                &       &H    &$3.14 \times 10^{-5}$ &$1.77 \times 10^{-4}$ &$0.0013$ \\
                &       &K    &$2.15 \times 10^{-6}$ &$1.21 \times 10^{-5}$ &$9.1 \times 10^{-5}$ \\
                &       &L$'$ &$4.76 \times 10^{-4}$ &$2.68 \times 10^{-3}$ &$0.02$ \\
                &       &M    &$2.31 \times 10^{-3}$ &$1.30 \times 10^{-2}$ &$0.097$ \\
\enddata

\tablenotetext{a}{Predicted secondary eclipse depth.  We have assumed
  a primary eclipse depth of $7.5\%$ based on the light curves shown
  in \S \ref{lc_sect}.}

\end{deluxetable}

For the purposes of these calculations, we have neglected stellar
insolation.  The zero albedo equilibrium temperature for NLTT~41135B
is approximately $460\ {\rm K}$, so it is likely that we underestimate
the surface brightnesses, especially at the two oldest ages where
insolation will represent a significant contribution to the brown
dwarf energy budget.

It is also important to consider the validity of the assumption of no
``reflected light'' from the brown dwarf, since although its
temperature is predicted to be very low, the radius should be only
$\simeq 1/2-1/3$ that of the primary.  For brown dwarfs with effective
temperatures $\la 1300\ {\rm K}$, complete gravitational settling of
any dust grains should occur (e.g. \citealt{allard2001}), which is the
physics assumed by the {\sc cond} models we have used, and should give
low albedo, since the atmosphere consists mostly of absorbing species.
We compute an additional set of predictions (shown in Table
\ref{secpredictreflect}) assuming an (unrealistically large) value for
the geometric albedo ($A_g$) equal to unity to illustrate an maximum
size of this effect\footnote{Upper limits from observations of
secondary eclipses and phase curves in the optical for Hot Jupiter
exoplanets, which are thought to have similar atmospheric
constituents, indicate that $A_g \la 0.1$ in the visible
(e.g. \citealt{rowe2008}).}, which are reproduced in Table
\ref{secpredictreflect}.  Comparing the results in the two tables
indicates that reflection should have only a very minor effect on our
predictions for the bandpasses where the secondary eclipse is most
likely to be observable.

\begin{deluxetable}{lrlrrr}
\tabletypesize{\normalsize}
\tablecaption{\label{secpredictreflect} Predicted luminosity ratios,
  surface brightness ratios, and secondary eclipse depths for
  NLTT~41135 with $A_g = 1.0$.}
\tablecolumns{6}

\tablehead{
\colhead{Age} & \colhead{$T_2$} & \colhead{Band} & \colhead{$L_2/L_1$} & \colhead{$J$} &\colhead{$d_2$\tablenotemark{a}} \\
& \colhead{(K)} & & & & \colhead{(\%)}
}

\startdata
$1\ {\rm Gyr}$  &$1067$ &z    &$1.22 \times 10^{-3}$ &$5.59 \times 10^{-3}$ &$0.042$ \\
                &       &J    &$3.87 \times 10^{-3}$ &$1.78 \times 10^{-2}$ &$0.13$ \\
                &       &H    &$2.33 \times 10^{-3}$ &$1.07 \times 10^{-2}$ &$0.08$ \\
                &       &K    &$1.95 \times 10^{-3}$ &$8.95 \times 10^{-3}$ &$0.067$ \\
                &       &L$'$ &$1.07 \times 10^{-2}$ &$4.92 \times 10^{-2}$ &$0.37$ \\
                &       &M    &$1.77 \times 10^{-2}$ &$8.11 \times 10^{-2}$ &$0.61$ \\
\hline
$5\ {\rm Gyr}$  &$643$  &z    &$4.27 \times 10^{-4}$ &$2.28 \times 10^{-3}$ &$0.017$ \\
                &       &J    &$4.94 \times 10^{-4}$ &$2.63 \times 10^{-3}$ &$0.02$ \\
                &       &H    &$4.19 \times 10^{-4}$ &$2.23 \times 10^{-3}$ &$0.017$ \\
                &       &K    &$3.30 \times 10^{-4}$ &$1.76 \times 10^{-3}$ &$0.013$ \\
                &       &L$'$ &$1.49 \times 10^{-3}$ &$7.96 \times 10^{-3}$ &$0.06$ \\
                &       &M    &$4.57 \times 10^{-3}$ &$2.44 \times 10^{-2}$ &$0.18$ \\
\hline
$10\ {\rm Gyr}$ &$533$  &z    &$3.40 \times 10^{-4}$ &$1.91 \times 10^{-3}$ &$0.014$ \\
                &       &J    &$3.45 \times 10^{-4}$ &$1.94 \times 10^{-3}$ &$0.015$ \\
                &       &H    &$3.25 \times 10^{-4}$ &$1.83 \times 10^{-3}$ &$0.014$ \\
                &       &K    &$2.96 \times 10^{-4}$ &$1.66 \times 10^{-3}$ &$0.012$ \\
                &       &L$'$ &$7.69 \times 10^{-4}$ &$4.33 \times 10^{-3}$ &$0.032$ \\
                &       &M    &$2.60 \times 10^{-3}$ &$1.46 \times 10^{-2}$ &$0.11$ \\
\enddata

\tablenotetext{a}{Predicted secondary eclipse depth.  We have assumed
  a primary eclipse depth of $7.5\%$ based on the light curves shown
  in \S \ref{lc_sect}.}

\end{deluxetable}

It is clear from the tables that the best prospects for detection of
the secondary eclipse lie at the longest wavelengths (L$'$ and M), and
will likely require space-based observations in order to obtain
adequate precision.  J and to some extent H show more favorable
eclipse depths than K due to the familiar blueward shift of the $J-K$
and $H-K$ colors for T-dwarfs resulting from H$_2$ and CH$_4$
absorption features.

Assuming M-band to be a reasonable proxy for the longest-wavelength
($4.5\ {\rm \mu m}$) {\it Warm Spitzer}\ channel, this eclipse might
be detectable, although it is likely that \spitzer's poor angular
resolution would result in the images of NLTT~41135 and 41136 being
blended, thus diluting the eclipse depths by a factor of $\simeq 3$.
Whilst the eclipses should still be detectable for the youngest ages we
have considered, this might become extremely challenging at the more
likely age of $\sim 10\ {\rm Gyr}$.  Due to the rapid decline in
J-band as a function of age, a similar statement applies for the
prospects of detection using near-infrared instruments aboard the {\it
  Hubble Space Telescope}.

With the {\it James Webb Space Telescope}, the secondary eclipse
detection should be straightforward, and the predicted angular
resolution is sufficient to resolve the M-dwarf pair even out to
the mid-infrared.

It is interesting to speculate that if the secondary eclipse was total
(which we stress is very unlikely given the essentially circular orbit
and grazing geometry of the primary eclipse), this would yield the
bandpass-integrated luminosity of the brown dwarf directly from the
eclipse depth, and would break the degeneracy inherent in the
modeling of the grazing eclipses in conjunction with the radial
velocity, yielding also the radii of both components.

Finally, we note that it may be possible to derive a relatively
crude constraint on the radius of the brown dwarf even in the most
likely case where both eclipses are grazing.  Recall from \S
\ref{lc_sect} that we assumed a prior on the system radius ratio in
order to determine $a/R_1$.  However if we knew the radius of the
M-dwarf, $a/R_1$ would be already completely determined and could be
fixed in the model, allowing us to instead constrain $R_2/R_1$.  This
could be done by measuring the angular diameter of NLTT~41135, in
conjunction with a parallax and a K-band magnitude measurement to
determine its mass.  Due to the remaining strong degeneracies between
$R_2/R_1$ and $i$ it is unlikely that this measurement would yield
$R_2/R_1$ to any better than $\pm 0.1$ (see Figure
\ref{lcmodparamsplot}).

\section{The NLTT~41135 / 41136 visual binary system}
\label{vb_sect}

Using the \citet{bcah98} models, we estimate a mass ratio of $\simeq
0.8$ for the M-dwarf visual pair from the effective temperatures in
Table \ref{photparams} (assuming ${\rm [Fe/H]} = -0.5$, $5\ {\rm
  Gyr}$ age and the value of $\beta$ from Table \ref{physparams}).
Assuming the mass of NLTT~41135 from Table \ref{physparams}, this
implies the mass of NLTT~41136 is approximately $0.21\ \msun$, and
the total system mass is therefore $M_{\rm tot} \simeq 0.37\ \msun$.
The angular separation of $2\farcs40$ corresponds to a projected
physical separation of $55\ {\rm AU}$ at $23\ {\rm pc}$ distance.
Given this information, we can estimate a minimum orbital period of
$680\ {\rm years}$ using Newton's version of Kepler's third law.

Given the long orbital period, it will be challenging to measure the
parameters of the astrometric orbit unless its orientation or
eccentricity are extremely favorable, although the motion of the
components relative to their presumed common proper motion could be as
large as a few tens of milliarcseconds per year especially if the
orbit is viewed face-on.  The maximum radial velocity difference
between the pair corresponding to this period is $2.4\ {\rm km\
  s^{-1}}$ (assuming an edge-on orbit), which is reasonably consistent
with our measurements from the TRES spectroscopy (see \S
\ref{tres_sect}) provided the measurements were taken at less than the
maximum separation or the orbit is not edge-on.

\section{Discussion}
\label{disc_sect}

We have reported the discovery of a hierarchical triple system
containing two M-dwarfs and a brown dwarf orbiting the less massive
of the pair in an eclipsing system.  The masses of the components are
approximately in the ratio $8:6:1$.

The existence of this object poses a challenge to theories of brown
dwarf formation.  The configuration of the system strongly resembles a
scaled-down (in mass) version of a relatively common configuration for
solar-type stars.  The mass ratio of only $6:1$ for the eclipsing
binary component of the system would seem to be difficult to produce
by disk fragmentation, because it would require an extraordinarily
massive disk surrounding the forming M-dwarf core.

Given this difficulty it seems likely that this object may have formed
in the same manner as its higher-mass analogs: directly from
gravitational collapse of an overdensity in a molecular cloud.
It is interesting to postulate that if many such systems are made by
the star formation process, there will be some that are dynamically
unstable, and may eject one or more of their components.  As proposed
in the ejection hypothesis of brown dwarf formation
\citep{reipurth2001}, this may offer a mechanism to produce single
brown dwarfs, although as discussed in \S \ref{intro_sect}, evidence
such as the existence of single brown dwarfs with discs argues against
this being the only mechanism for brown dwarf formation.

It is interesting to compare the secondary of the present system with
the secondary of the CoRoT-3 system, a $22\ \mjup$ brown dwarf
orbiting an F3V star \citep{deleuil2008}.  The brown dwarfs in the two
systems have similar masses, and their orbital periods differ by only
$50\%$, yet CoRoT-3b orbits a star of luminosity $10^3$ times greater
than does NLTT~41135B, and thus experiences a $500$ times greater
(bolometric) stellar insolation.  This means that while the energy
budget of NLTT~41135B is most likely dominated by internal energy
sources, CoRoT-3b will be dominated by insolation.  It would thus be
extremely interesting to compare the observed physical properties of
these objects to obtain insights into brown dwarf physics, for example
the efficiency of re-distribution of energy from the day-side to the
night-side of CoRoT-3b, as has been done for extrasolar planets
(e.g. \citealt{knutson2007}).  We also note that these objects may
have formed through different mechanisms, since the mass ratio for the
CoRoT-3 system of $q = 0.015$ (an order of magnitude lower than for
NLTT~41135B) may be difficult to produce via conventional mechanisms
thought to be responsible for binary star formation.

NLTT~41135B is also of comparable mass to the secondary in the
double-lined brown dwarf EB 2MASS~J05352184-0546085
(\citealt{stassun2006,stassun2007}), although it is of much older age
and orbits a main sequence star.  Nonetheless, these objects are
probably quite close to being evolutionary analogues of one another,
and both have an energy budget most likely dominated by internal
energy sources.

The present system provides a dynamical mass for a field brown dwarf,
and offers the potential to allow models of brown dwarf evolution to
be tested at extremely old ages, where dynamical constraints are
scarce.  In order to test the theory, we must attempt to determine or
constrain the system age, in addition to measuring another fundamental
property of the brown dwarf.  The most accessible measurement is the
secondary eclipse, which in conjunction with the primary eclipse
yields a measurement of the central surface brightness ratio of the
two components from the ratio of the eclipse depths.

Finally, we note that a major limitation in the determination of the
brown dwarf mass and M-dwarf parameters is the uncertainty in the mass
of NLTT~41135A.  We suggest that the best prospect for improvement
would be to obtain an accurate trigonometric parallax for this object,
in conjunction with resolved near-infrared apparent magnitude
measurements (preferably in K-band).  This should yield an estimate
for the metallicity, and thus the primary mass to $\sim 10\%$, giving
an improved solution for the eclipsing binary to more tightly
constrain models of low-mass stars and brown dwarfs.

\acknowledgments The MEarth team gratefully acknowledges funding from
the David and Lucile Packard Fellowship for Science and Engineering
(awarded to DC).  This material is based upon work supported by the
National Science Foundation under grant number AST-0807690.  LB and
DWL acknowledge partial support from the NASA Kepler mission under
cooperative agreement NCC2-1390.  JAJ thanks the NSF Astronomy and
Astrophysics Postdoctoral Fellowship program for support in the years
leading to completion of this work, and acknowledges support from NSF
grant AST-0702821.  We thank Isabelle Baraffe for providing {\sc
  NextGen} and {\sc cond} models in $z$-band, Daniel Fabrycky for
helpful discussions regarding dynamics, and Timothy Brown and the rest
of the team at the Las Cumbres Observatory Global Telescope for their
efforts in trying to obtain a resolved light curve of the system.  The
MEarth team is greatly indebted to the staff at the Fred Lawrence
Whipple Observatory for their efforts in construction and maintenance
of the facility, and would like to explicitly thank Wayne Peters, Ted
Groner, Karen Erdman-Myres, Grace Alegria, Rodger Harris, Bob
Hutchins, Dave Martina, Dennis Jankovsky and Tom Welsh for their
support.  Finally, we thank the referee for a thorough and helpful
report, which has substantially improved the manuscript.

Based on observations made with the Nordic Optical Telescope, operated
on the island of La Palma jointly by Denmark, Finland, Iceland,
Norway, and Sweden, in the Spanish Observatorio del Roque de los
Muchachos of the Instituto de Astrofisica de Canarias.
This research has made extensive use of data products from the Two
Micron All Sky Survey, which is a joint project of the University of
Massachusetts and the Infrared Processing and Analysis Center /
California Institute of Technology, funded by NASA and the NSF, NASA's
Astrophysics Data System (ADS), and the SIMBAD database, operated at
CDS, Strasbourg, France.  The Digitized Sky Surveys were produced at
the Space Telescope Science Institute under U.S. Government grant NAG
W-2166. The images of these surveys are based on photographic data
obtained using the Oschin Schmidt Telescope on Palomar Mountain and
the UK Schmidt Telescope. The plates were processed into the present
compressed digital form with the permission of these institutions.

Funding for the SDSS and SDSS-II has been provided by the Alfred
P. Sloan Foundation, the Participating Institutions, the National
Science Foundation, the U.S. Department of Energy, the National
Aeronautics and Space Administration, the Japanese Monbukagakusho, the
Max Planck Society, and the Higher Education Funding Council for
England. The SDSS Web Site is {\tt http://www.sdss.org/}.
The SDSS is managed by the Astrophysical Research Consortium for the
Participating Institutions. The Participating Institutions are the
American Museum of Natural History, Astrophysical Institute Potsdam,
University of Basel, University of Cambridge, Case Western Reserve
University, University of Chicago, Drexel University, Fermilab, the
Institute for Advanced Study, the Japan Participation Group, Johns
Hopkins University, the Joint Institute for Nuclear Astrophysics, the
Kavli Institute for Particle Astrophysics and Cosmology, the Korean
Scientist Group, the Chinese Academy of Sciences (LAMOST), Los Alamos
National Laboratory, the Max-Planck-Institute for Astronomy (MPIA),
the Max-Planck-Institute for Astrophysics (MPA), New Mexico State
University, Ohio State University, University of Pittsburgh,
University of Portsmouth, Princeton University, the United States
Naval Observatory, and the University of Washington.

The authors wish to recognize and acknowledge the very significant
cultural role and reverence that the summit of Mauna Kea has always
had within the indigenous Hawaiian community.  We are most fortunate
to have the opportunity to conduct observations from this mountain.

{\it Facilities:} \facility{UH:2.2m (OPTIC)}, \facility{FLWO:1.2m (KeplerCam)}, \facility{FLWO:1.5m (TRES)}, \facility{NOT (FIES)}

\end{document}